# Reference-Based Publication Networks with Episodic Memories

Anthony F.J. van Raan[1]
Centre for Science and Technology Studies
Leiden University
Wassenaarseweg 52, P.O.Box 9555
2300 RB Leiden, The Netherlands

*In this paper we report first results of our study on network characteristics of a reference-based, bibliographically coupled (BC) publication network structure. We find that this network of clustered publications shows different topologies depending on the age of the references used for building the network. A remarkable finding is that only the network structure based on all references within publications is characterized by a degree distribution with a power-law dependence. This topology, which is typical for scale-free networks, disappears when selecting references of a specific age for the clustering process. Structuring the publication network as a function of reference age, allows 'tuning through the episodic memory' of the nodes of the network. We find that the older the references, the more the network tends to change its topology towards a Gaussian degree distribution.*

## 1. Introduction

References are important characteristics of a publication. We studied linkages and clustering of publications from the year 2001 with help of 'bibliographic coupling' (BC) on the basis of their references (i.e., citations given to earlier publications) and measured the characteristics of the emerging network structure.

In bibliographic coupling, two articles are linked if they have at least one reference in common. Thus, a larger part of the scientific literature is structured by a network of interlinked publications that are often grouped in clusters. The BC structure is a rather unorthodox type of citation-based publication network in which *recent* literature --in this case publications of 2001-- is structured in terms of clusters on the basis of co-referencing, whereas in co-citation analysis (CC) *older* literature, namely references in 2001-publications are structured in terms of clusters on the basis of their co-citing papers. In other words, in BC the 'nowadays landscape' of scientific literature is created on the basis of their memories to older literature, and in CC a landscape of 'older literature' is created reflected as it were from nowadays publications.

In the usual citation networks studied so far (e.g., Redner 1998 [1]; Vasquez 2001 [2]; Klemm and Eguíluz 2002a [3]; Newman 2003 [4]) the nodes (or: vertices) are published articles and a directed link (or: edge) from article A to a previously published article B indicates that A cites B, i.e., article A gives a reference to article B. Measurement of the number of times a publication (node of the network) is cited, yields the 'incoming' degree distribution (Dorogovtsev and Mendes 2002 [5]). Thus, the degree distribution P(k) gives the probability that a randomly selected node has k links. The degree distribution is a kind of stationary (a time-independent) measure of the network (Barabási *et al* 2002 [6]).

---

[1] *E-mail:* vanraan@cwts.leidenuniv.nl

In a randomly wired network (random graph model) the nodes have a uniformly distributed probability to connect. The probability that a node has k links, and with that the degree distribution of these random networks, then follows a Poisson-distribution. An important characteristic of real networks, however, is local clustering which means that network-structures are more complex than simple randomly wired networks. It appears that many real networks are scale-free, i.e., their degree distribution follows a power law. Large networks may self-organize into such a scale-free state. Scale-free means that a functional form f($x$) remains unchanged under rescaling of a variable $x$, which means f(a$x$) = bf($x$). The solutions to this general equation are always power law forms.

The functionality of the network heavily depends on the type of distribution. Hence, characteristics such as degree distributions are not just of statistical interest. These distributions describe the topology of networks, and topology is directly related to important features of a network such as signal-propagation speed (Watts and Strogatz 1998 [7]). An extensive overview of the statistical mechanics of complex networks is given by Albert and Barabási (2002) [8].

Our study aims at finding statistical properties, such as degree distribution, path lengths, connectivity distributions and dynamical aspects that characterize the structure and behavior of BC publication networks. In this paper we focus on phenomena related to the degree distributions. We also aim to understanding the meaning of these properties.

We take an analogy with scientific collaboration networks. In scientific collaboration networks, two nodes (authors) are linked if they coauthored one or more publications. These 'co-author' networks recently represent a kind of archetype example of a complex evolving network (Newman 2001a [9], 2001b [10], 2001c [11] on static properties; and Barabási *et al* 2002 [6] on dynamical properties). Therefore, we illustrate the analogy of our BC publication network with the scientific collaboration network in the following scheme:

___

| | |
|---|---|
| Authors have publications | Publications have references |
| Publications may have more than one author | References may appear in more than one publication |
| These authors are called co-authors | These publications are called bibliographically coupled (BC) publications |
| An author may have *s* co-authors | A reference may have *s* BC publications |
| We call this a collaboration cluster of size *s* | We call this a BC publication cluster, size *s* |
| In a scientific collaboration network, authors are the nodes and co-authorship establish the links | In a scientific BC publications network, publications are the nodes and bibliographic coupling establish the links |
| Number of publications per author | Number of references per publication |
| Number of authors per publication | Number of publications per reference (this means in fact the number of citing papers in a year to a specific reference) |

There is an interesting difference between the two above networks. In the scientific collaboration network, or 'co-author' network, a publication functions as an 'affinity characteristic' of authors: it is the element which causes scientists to cluster as co-authors. Publications (the 'clustering elements') and authors (the 'clustered elements') are, however, completely different entities in nature. In the BC publication network, a reference functions as an 'affinity characteristic' of publications: it is the element which causes publications to cluster as BC co-publications. But references are publications themselves, so in our network the 'clustering elements' and the 'clustered elements' are the same things in nature. It is like clustering people on the basis of their parents, grandparents, and further forefathers.

In the scientific collaboration network the co-authors know each other, they form a social structure of personal relations. In the BC publication network, the co-publications 'do not know each other', they just share one or more references, and in that sense this network looks like a large consumer system in which we find clusters of consumers (who do not necessarily know each other) sharing an interest in a specific groups of products or of services. In this study, the 'shared interest' is a research theme or research area.

## 2. Basic principles of BC publication networks

Our network consists of linked 2001-publications. These publications are connected by their referencing characteristics. This type of clustering is called *bibliographic coupling* (BC) as opposed to *co-citation coupling* (CC). The history of co-citation study bibliographic coupling studies goes back to Kessler in the early 1960's. An extensive overview of the work of Kessler is given by De Solla Price in his pioneering work 'Networks of Scientific Papers' (Price 1965 [12]).

We first explain the main lines of our method. We define a publication as a function of its references. As a simple example we take a small publication data set in which we have four publications $p_1$, $p_2$, $p_3$ and $p_4$, and 5 references $r_1$, $r_2$, $r_3$, $r_4$, and $r_5$. In bibliometric language, the $p_i$ are the *citing* papers, and the $r_i$ the *cited* papers. Say $p_1$ contains all references $r_1$ to $r_5$; $p_2$ contains $r_1$, $r_3$ and $r_4$; $p_3$ has only $r_1$ and $r_4$ as a reference, and $p_4$ has none of the five references in its reference list. We can now construct a publication-to-reference matrix $\underline{P}$:

|       | $r_1$ | $r_2$ | $r_3$ | $r_4$ | $r_5$ |
|-------|-------|-------|-------|-------|-------|
| $p_1$ | 1     | 1     | 1     | 1     | 1     |
| $p_2$ | 1     | 0     | 1     | 1     | 0     |
| $p_3$ | 1     | 0     | 0     | 1     | 0     |
| $p_4$ | 0     | 0     | 0     | 0     | 0     |

We observe that $p_1$ is bibliographically coupled to $p_2$ via $r_1$ (and also via $r_3$ and $r_4$, but one 'link' is sufficient to have a bibliographic coupling). It is clear however that this number of links --three in this case-- can be used as a measure of strength of the bibliographic coupling between two articles. Also, $p_1$ is coupled to $p_3$ (via $r_1$, or via $r_4$), but not to $p_4$. Thus, $p_1$, $p_2$ and $p_3$ form a BC-cluster, in which $p_1$ has two BC 'co-publications' (and the same is true for $p_2$ and $p_3$). Notice that $p_1$, $p_2$, $p_3$ and $p_4$ have together *10 references*, which, however, represent *5 different cited articles*.

Pre-multiplication of matrix $\underline{P}$ with its transpose $\underline{P}^T$ yields the (symmetric) reference-correlation matrix:

$$\underline{C} = \underline{P}^T * \underline{P} \qquad (1)$$

which is in our example:

| **3** | 1     | 2     | *3*   | 1     |
|-------|-------|-------|-------|-------|
| 1     | **1** | 1     | 1     | 1     |
| 2     | 1     | **2** | 2     | 1     |
| *3*   | 1     | 2     | **3** | 1     |
| 1     | 1     | 1     | 1     | **1** |

The *diagonal* values (printed in bold face) of this matrix $\underline{C}$, $c(i,i) = c(i)$, indicate the number of times a specific cited publication (reference) is mentioned in the total set of publications. This is 3 times for $r_1$, 1 time for $r_2$, and so on. Thus, the matrix diagonal represents the *occurrences* of each reference, or, in other words, the number of citations given by the set of publications to each cited publication. This means that the distribution function of $c(i)$ gives the *in-degree* distribution of the publication set: the number of (citing) publications per cited publication (reference). We discuss this in-degree distribution $N(c)$ for our empirical data in the next section.

The *off-diagonal* values give the *co-occurrences*, for instance $r_1$ and $r_4$ (value printed in italics) are mentioned 3 times together in publications of the set, namely in $p_1$, $p_2$, and $p_3$, but for this information we have to go back to the original matrix $\underline{P}$ as in matrix $\underline{C}$ all 'direct' information on the *publications* is 'lost'.

These co-occurrences of references are the basis of what is called in bibliometric studies '*co-citation* linkage clustering' (CC). For instance: $r_1$ and $r_4$ are linked with 'strength' 3, but $r_4$ is also linked to $r_3$ with strength 2, and so on.

If we now take the mirrored matrix multiplication of Eq. 1, i.e., post-multiplication of our original matrix with its transpose, we get the (symmetric) publication-correlation matrix:

$$\underline{R} = \underline{P}(r) * \underline{P}^T(r) \qquad (2)$$

which is in our example:

```
5   3   2   0
3   3   2   0
2   2   2   0
0   0   0   0
```

The *diagonal* values (again printed in bold face) of this matrix $\underline{R}$, $r(i,i) = r(i)$, indicate the number of references in each publication. This is 5 for $p_1$, 3 for $p_2$, and so on. This means that the distribution function of $r(i)$ gives the *out-degree* distribution of the publication set: the number of references (cited publications) per (citing) publication. We discuss this out-degree distribution $N(r)$ for our empirical data also in the next section.

The *off-diagonal* values give the number of references shared by any two publications, for instance $p_1$ and $p_3$ (value printed in italics) share 2 references (namely in $r_1$ and $r_4$, but also for this information we have to go back to the original matrix $\underline{P}$ as in matrix $\underline{R}$ all 'direct' information on the *references* is 'lost'). Thus, $\underline{R}$ provides the *strengths* of the links between each possible publication pair within the set and can be used for calculating 'distances' between publications.

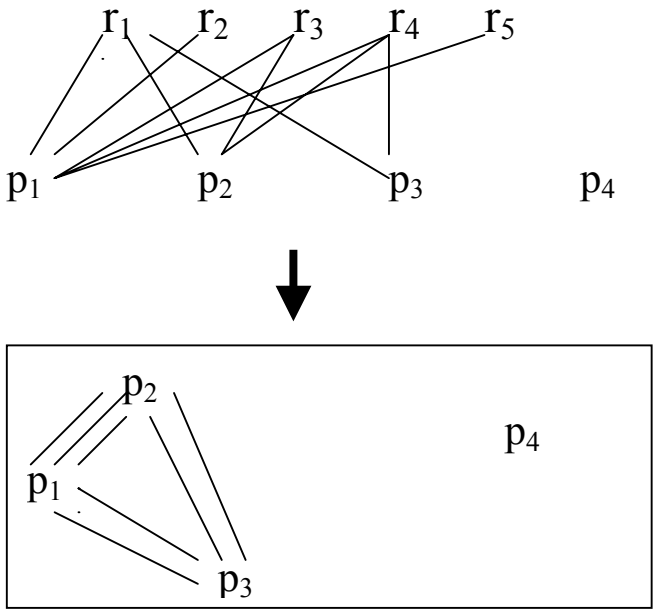

*Fig. 1: Graph of the BC network example*

The structure of the BC-coupled publications can be represented in a bipartite graph, and the resulting network as a simple projection of that graph, as given in Figure 1, together with the strengths of the links as follow from matrix ***R***. We observe that $\sum_{i,j(i\neq j)} r_{ij}$ gives the total number of links to node (publication) $p_i$, for instance for $p_1$ this number is 5, for $p_2$ it is also 5, and for $p_3$ it is 4. The above discussion shows that co-citation linking and bibliographic coupling are mathematically related by simple matrix algebra.

## 3. Results and Discussion

### 3.1 Characteristics of the BC-network

We analyzed all 2001 publications in the complete set of citation indexes[2], which totals to 1,099,017. The following characteristics of the entire data set were studied.

*1. Number of references per publication*

The number of references per publication is an out-degree measure (analogous to the number of publications per author in the case of scientific collaboration), i.e., the distribution N($r$) of 2001-publications as a function of the number of references, see Figure 2. For instance, we find that there are 43,364 publications in 2001 having just 1 reference, and one publication having 2,301 references.

This distribution function N($r$) appears to consist of two power-law regimes. The tail, i.e., the higher-$r$ part (from about $r = 40$) of the distribution, follows a steep power-law. This is the part of the distribution where a considerable amount of the publications are review articles. We find a power-law decay with exponent approximately – 3.7. The low-$r$ part (particularly for $r$ between 1 and 10) is quite flat and follows a very slowly decreasing power-law. Here most of the publications will be shorter articles such as letters, notes. Apparently there is not much difference in the (relatively small) number of references for these types of publications. Taking into account only those references that are themselves articles covered by the ISI databases and published after 1980 (in order to get a pair-wise match of a reference with a source article in our database), we find a power-law decay of approximately –3.5.

In his seminal paper, De Solla Price (Price 1965 [12]) reports a power-law behavior in the tail of the distribution with an exponent approximately –2.0 for the out-degree of the citation network ('incidence of references'). He used the references of papers published in 1961. Also the flat low-$r$ part was observed. It is not yet clear why we find a much steeper decay than in the work of De Solla Price. Possibly reference characteristics of publications have been changed in the last 40 years since Price's observations. Therefore, we are currently investigating the number of references per publication as a function of publication year with the data available in our bibliometric data-system, ranging from 1980 up till now.

---

[2] The Science Citation Index (SCI), the Social Science Citation Index (SSCI), the Arts & Humanities Citation Index (AHCI), and all 'specialty' indexes such as Neurosciences, Biochemistry and Biotechnology, etc., published by the Institute of Scientific Information (ISI) in Philadelphia.

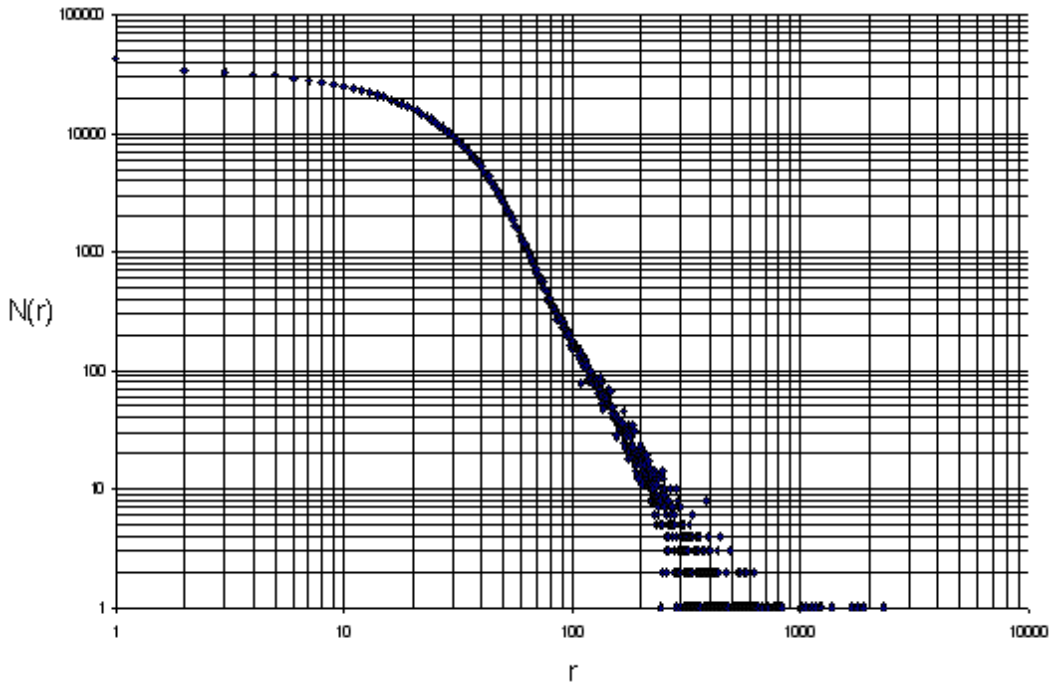

*Fig. 2: Number of references per publication (2001)*

Using the analogy of our BC-network with the scientific collaboration network, we may compare the number of references per publication with the number of publications per author. For this latter distribution, Newman (2001a [9]) finds an exponentially truncated power law. He attributes this truncation to the finite time window of five years used in his study, which prevents authors from having a very large number of papers. For instance, the co-author degree distribution for the Los Alamos Archive can be described with a truncated power law P(k) = C $k^{-\tau}$ exp(-k/κ), τ and κ are specific parameters, see Newman (2001a) [9]. In the original work of Lotka (1926) [13], which shows a 'complete' power-law, a more 'life time' approach was taken and therefore such a window was not used. Alternative explanations for the truncation are based on specific growth models of networks (Krapivsky *et al* 2000 [14]; Barabási *et al* 2002 [6]) or specific collaboration models (Albert and Barabási 2000 [15]).

*2. Number of citations to the references*

Number of citations in 2001 (i.e., number of citing publications in 2001) to references, N(*c*), is an *in*-degree measure, analogous to the number of authors per paper in the case of scientific collaboration. Thus, we take all references in the 2001-publications, and analyze the number of times these references are cited by the 2001-publications. The results are shown in Figure 3. In 2001 15,301,841 references are given and these references represent in fact 4,876,752 'unique' references (i.e., cited articles). We find 2,230,575 *cited articles* that are cited only once in 2001-publications; 203,407 articles cited five times in 2001; 401 articles cited fifty times in 2001, and 1 article that is cited 3,484 times. This is, by definition, the most cited paper (from 1980 on) in 2001.

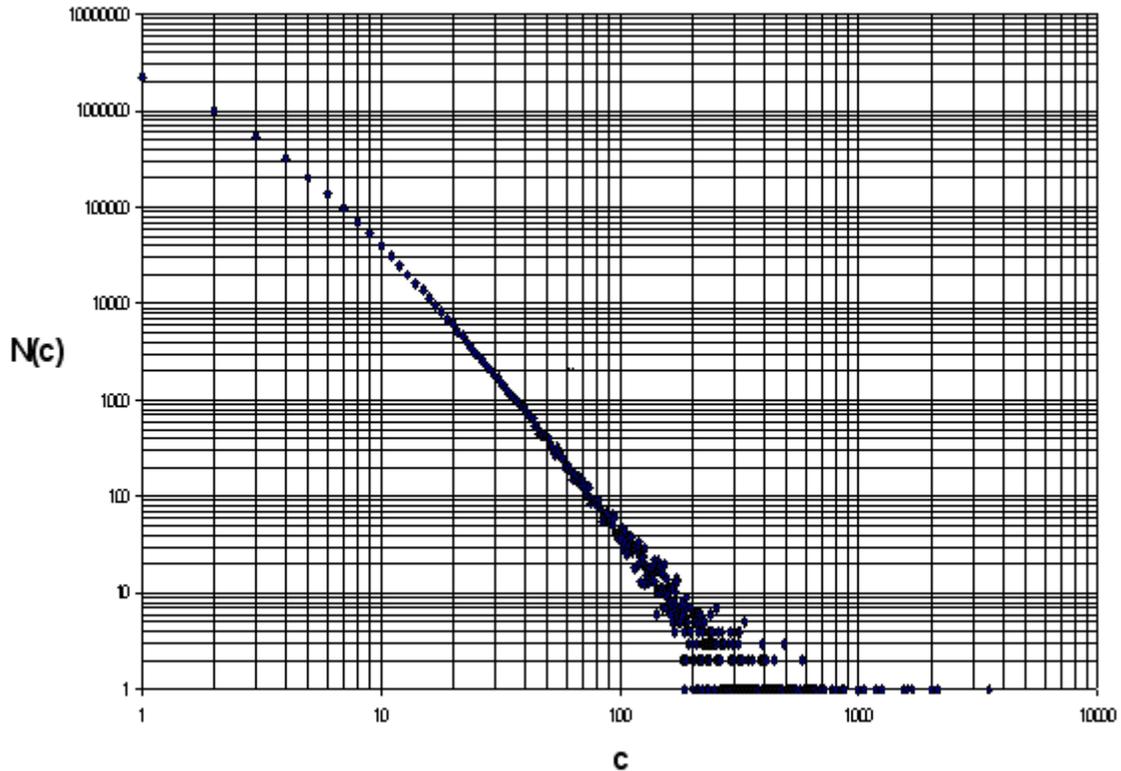

*Fig. 3: Number of citations to references of 2001-publications*

The citation distribution function shows for the major part a power-law decrease with a cut-off for lower *c*-values (*c* from 1 to about 10). We find a power-law decay with exponent approximately $-3.1$.

This power-law distribution for citations is well-known. In the above mentioned pioneering work, De Solla Price also studied the in-degree distribution ('incidence of citations') and reports a power-law distribution with an exponent between about $-2.5$ and $-3.0$ (Price 1965 [12], 1976 [16]). Also Naranan (1971) [17], using a subset of the data of Price, finds a power law exponent close to $-3.0$.

On the basis of his observations, De Solla Price developed the model of 'cumulative advantage', building on Simon's work on the Matthew effect, i.e., the rich get richer (see for instance Bornholdt and Ebel 2001 [18] and the original work of Simon 1955 [19]). In network-language, this phenomenon is a striking example of 'preferential attachment' (i.e., the probability for a node to obtain a new link increases with the number of links this node already has) as in citation networks a new publication is likely to cite a well-known and thus mostly much-cited publication more than a less cited publication (Barabási and Albert 1999 [20]; Dorogovtsev and Mendes 2002 [5]).

Measuring citation distributions are not so straightforward as often thought. There are quite different modalities of measurement. Redner (1998) [1] analysed citations to 1981-publications received in the years 1981-1997, thus the in-degree of publications of one year (1981). This is the case of a *fixed publication-year* (1981) followed by a wide 'window' of citation years. In our study we have a *fixed citation-year* (2001) and a wide window of preceding publications years. Also De Solla Price used a fixed (1961) citation year. Redner finds in his study that the asymptotic tail of the citation distribution appears to be described by a power law with exponent approximately –3. But given the quite different behaviour of his distribution function for high versus low citation numbers, he suggests two different 'citation regimes', in the sense that there might be different underlying mechanisms and thus different statistical features between less-cited (exponential behaviour) and highly-cited papers (power law). We think, however, that further studies are necessary to investigate the effects caused by the difference in measuring modalities, as in our case the deviation from a power law behaviour for low citation number is less stronger than in the study of Redner.

Laherrère and Sornette (1998) [21] and also Tsallis and De Albuquerque (2000) [22] state that natural phenomena often exhibit a power-law followed by a significant curvature. They question whether these observed deviations form a power-law behavior just simply result from finite-size effects or the existence of two regimes that are different in nature. They discuss models in which the distribution of citations of scientific papers can be fitted over the entire range of citation numbers with one single curve, so called stretched exponentials $n(x) \sim \exp\{-[(x/x_0)^\beta]\}$ (for the specific parameters $x_0$ and $\beta$ see Tsallis and de Albuquerque, 2000 [22]). Laherrère and Sornette (1998) [21] apply these stretched exponentials (yielding 'parabolic fractals') to citations of highly cited physicists. Other examples are the size-distribution of cities. This is important, as in many cases claimed power laws clearly show a 'parabolic' effect for high k-values instead of a 'real straight line' (in a log-log plot), see for instance Fig. 2b and also Figs. 14 and 15 in Barabási et al (2002) [6]. In these latter figures we see that this 'parabolic' effect is stronger with less nodes.

In a recent paper (van Raan 2001 [23]) we studied the in-degree distribution of about 15,000 chemistry publications published in The Netherlands in the period 1985-1993. Citations were counted in a modality again different from the two earlier mentioned. It is the modality often used in bibliometric analysis for evaluation purposes. For each publication year within the range 1985-1993, a 3-year window to receive citations after the year of publication is used. For instance, for publication year 1985 the citation window is 1986-1988, and so on. This measurement modality has the advantage of giving each publication year the same time period for receiving citations.

The resulting distribution function shows, for the larger number of citations, approximately a power law with an exponent of about –2.6. We observe similar 'parabolic' deviations from the 'ideal' power law as discussed above: an inclination to 'saturate' for the lower citation values, as well as a cut-off for higher citation values. In contrast to fitting procedures as in the work of Laherrère and Sornette (1998) [21] and of Tsallis and De Albuquerque (2000) [22], we developed a novel, *ab initio* theoretical model for the acquisition of citations by publications on the basis of a two-step competition process. Surprisingly, the result of this model is not the prediction of a power law behavior for the citation distribution. We find a second order Bessel function. And even more surprisingly, this second order Bessel function describes the empirically measured distribution function very well, for the entire range of citation values. This would mean, that the mechanism of citation distribution only 'mimics' a scale-free (power law) behavior. We are currently investigating the ability of our model to describe the in-degree distribution of the citation data in this study. A short presentation of our approach is given in the appendix, we refer for details and comparison with empirical results to Van Raan (2001) [23].

*3. Number of bibliographically coupled publications per publication*

The number of bibliographically coupled publications, or 'BC co-publications', per publication, N(*s*), is the main characteristic of our reference-based publication network system. Using again our analogy with the case of scientific collaboration, it is comparable with the number of coauthors (or: collaborators) per authors (see the scheme in Section1). A group of BC co-publications can be considered as a cluster of publications (just as a group of coauthors can be considered as a cluster), and therefore we also use the term 'cluster size' when dealing with the number of bibliographically coupled publications.

The distribution of the number of BC co-publications (BC-cluster size distribution) *based on all references of the 2001-publications* is presented in Figure 4a.

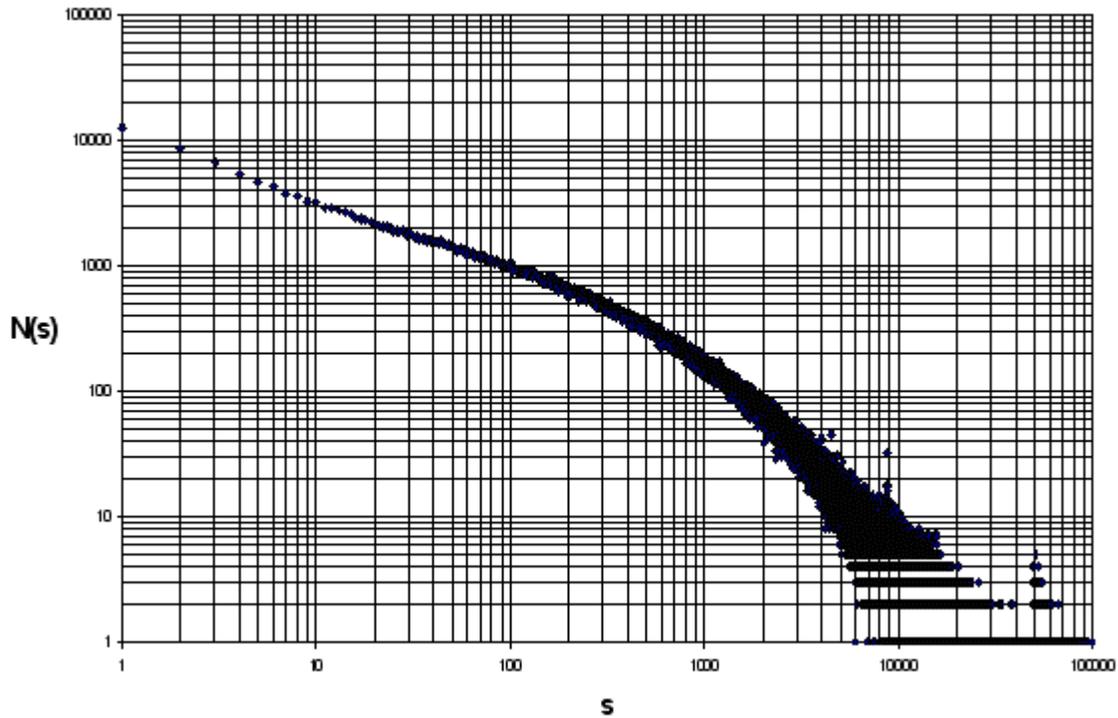

*Fig. 4a: Number of BC clusters, based on all references*

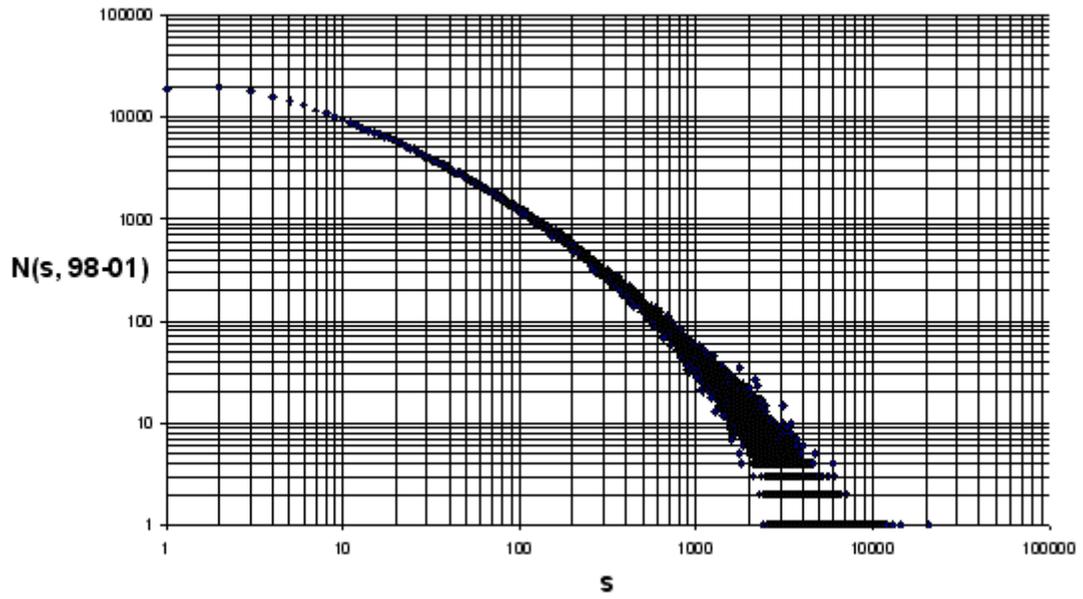

*Fig. 4b: Number of BC clusters, based on 1998-2001 references*

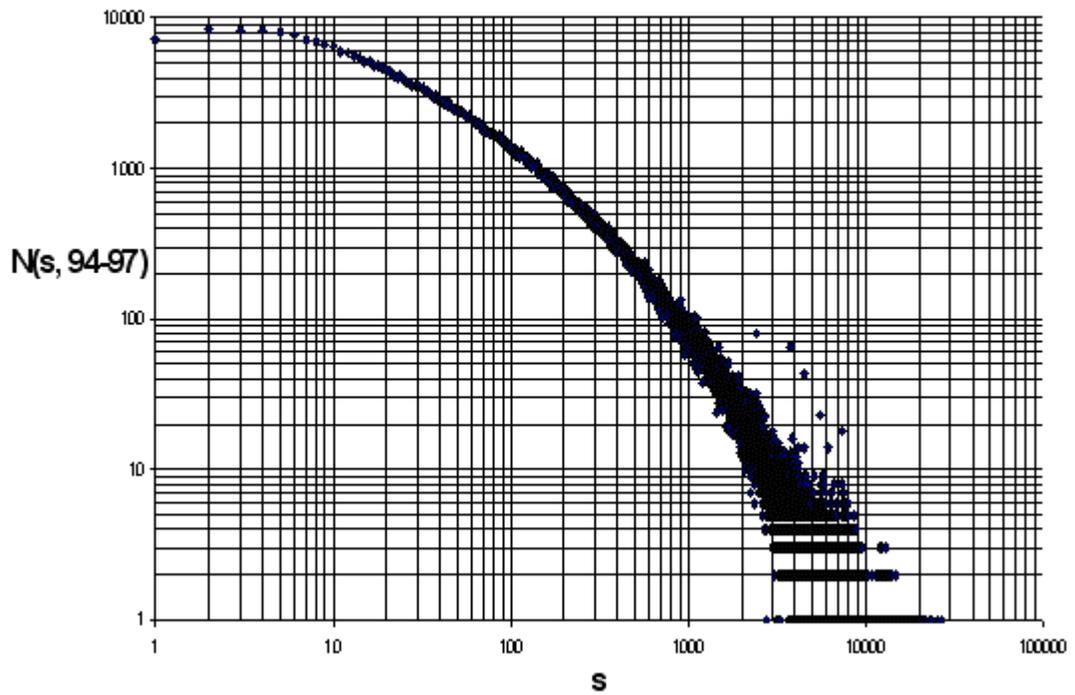

*Fig. 4c: Number of BC clusters, based on 1994-1997 references*

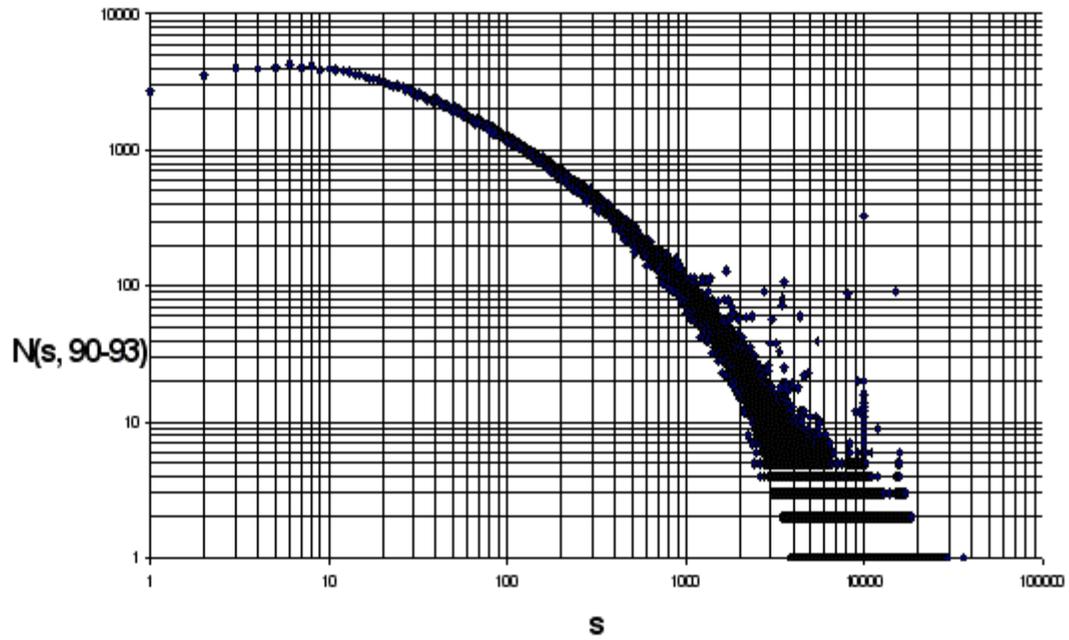

*Fig. 4d: Number of BC clusters, based on 1990-1993 references*

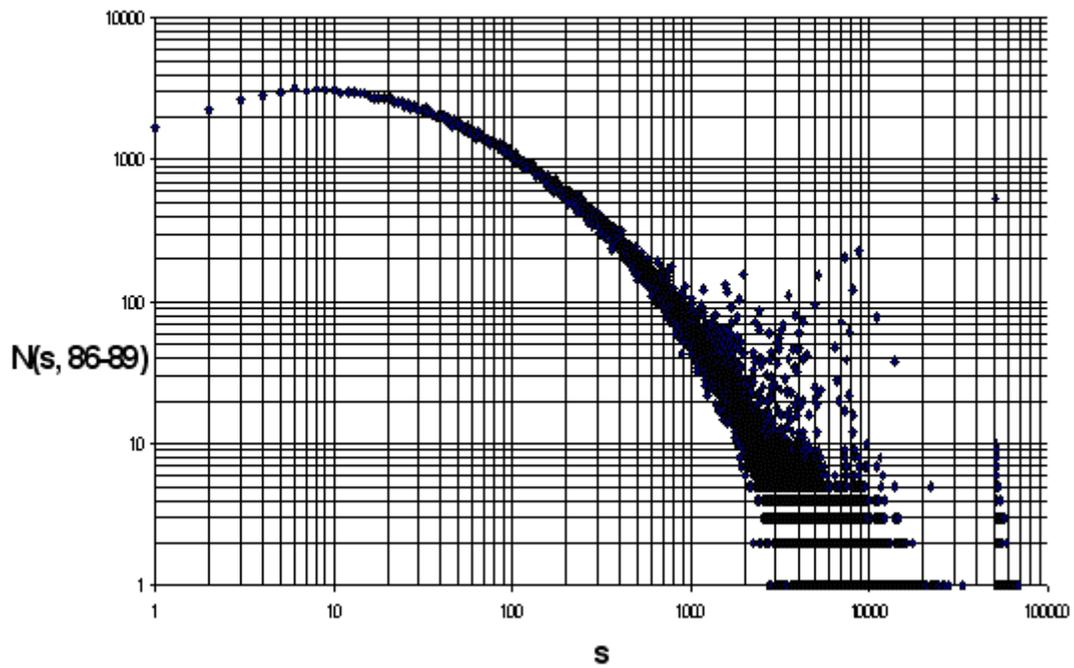

*Fig. 4e: Number of BC clusters, based on 1986-1989 references*

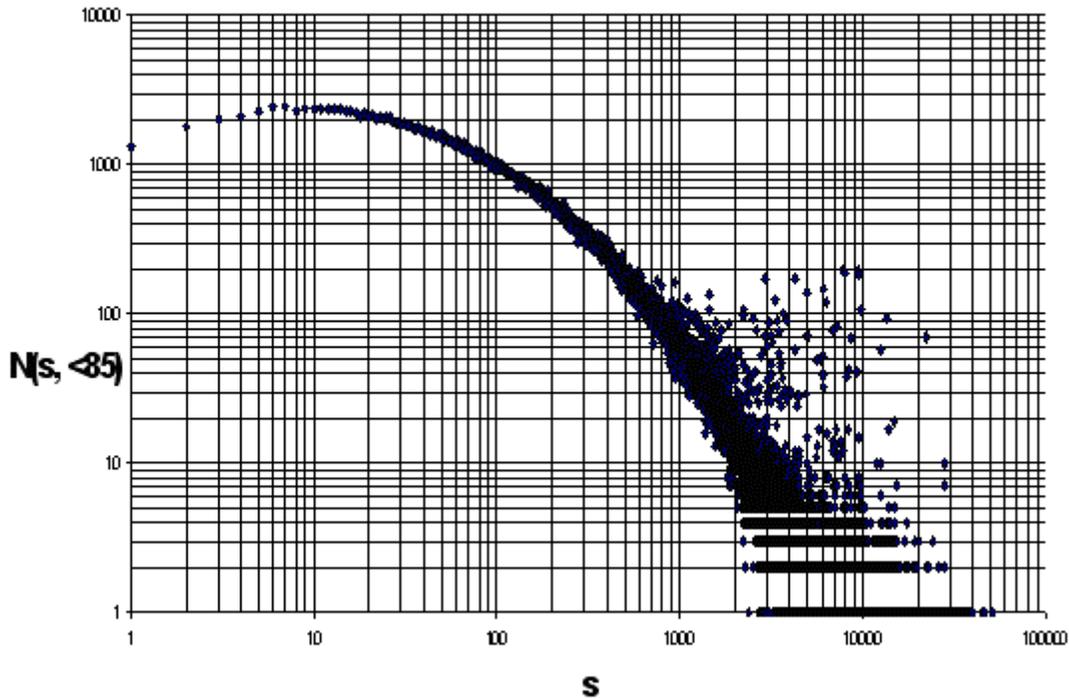

*Fig. 4f: Number of BC clusters, based on 1985 and earlier references*

In order to investigate an 'episodic memory effect' in the emergence of BC clusters, we also studied the distribution as a function of the age of the references, see Figs. 4b, c, d, e, and f. This age-dependent measurement is performed by selecting from the total set of references five different subsets: the references (given in 2001-publications) with publication years 1998-2001 (the 'youngest' references, Fig. 4b), publication years 1994-1997 (Fig. 4c), publication years 1990-1993 (Fig. 4d), publication years 1986-1989 (Fig. 4e), and, finally, references with publication years 1985 and before (the 'oldest' references, Fig. 4f), respectively.

The analyses show remarkable results. In the case where publications are characterized by *all* their references (Fig. 4a), we find a BC-cluster size distribution that is typical for a 'scale-free' network, i.e., with a power-law behaviour, with an exponential cut-off for cluster sizes above about 1,000. The *older* the references used to construct the BC network, the *stronger* the deviation of the distribution from a power-law toward a more exponential behaviour. This would mean that publications characterized by just their oldest references, cluster in a much more random way than if the entire list of references is taken into account. In other words, clusters based on 'old memories' tend to be distributed more 'normally', which also means less small clusters, as can be observed in Figs. 4b-f. As soon as the nodes 'rejuvenate', i.e., increase their 'short term memory', they tend to form a more power-law structured, i.e., scale-free network (much more small clusters). This tendency to deviate from a power law toward a more exponential behaviour in networks with 'aging of sites' is also observed in the model of Dorogovtsev and Mendes (2000) [24].

How can we explain this? The relatively old references are 'archival' and mostly much more *general* or 'classic' than the more recent references, which are typical field- or research theme-*specific*. Thus, these older references tend to link more parent publications, and the wiring of the BC network will therefore be more randomly distributed among the participating nodes, i.e., those 2001-publications having these relatively old references. Barabási *et al* (1999) [25] show that in case of a growing network the degree distribution function has an exponential form in case of 'uniform attachment', i.e., the new node connects with equal probability to the nodes already present in the system, independent of the degree values of a node (no preferential attachment, their model B, see also Albert and Barabási 2002 [8]). In our 'tuning' through the references, we more or less simulate a similar process in an otherwise static structure.

Smaller clusters are typical for research on very specific themes and in most cases these themes are very recent and thus characterized by relatively young references (the 'short-term memory of the system'). We indeed observe much more smaller clusters on the basis of 1998-2001 references (order of magnitude: 10,000) than in the cases of older references (order of magnitude: 1,000).

There is, however, an analytical problem. Since the *number* of references given by the citing publications is age-dependent, selection of increasingly older references also implies a choice for *increasingly less* references to characterize a publication. The largest group of references in publications concerns the most recent references, i.e., references to papers from 1998-2001. Also, it is obvious that the more references are included in the clustering process, the more larger clusters will be found. Indeed, we observe that the distribution based on the entire reference lists (Fig. 4a) contains the largest clusters.

In order to distinguish between 'time-dependent' and 'less references', we again performed the BC wiring process, but now with removing randomly 10% of the references. As an example of this data manipulation for the BC clustering process with the oldest references (i.e., from 1985 an earlier) is given in Fig. 5. We immediately observe that is no significant difference in the shape of the distribution as compared with the 'complete' data shown in Fig. 4f. Thus, we conclude that the topological changes reported in this paper are indeed due to time-dependent effects.

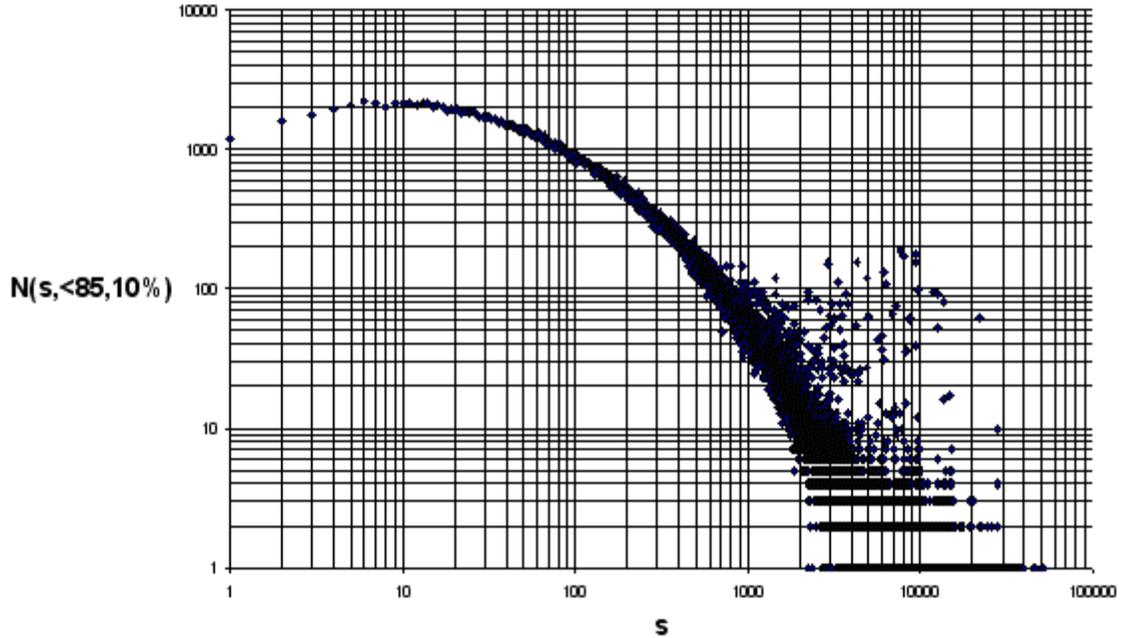

*Fig. 5: Number of BC clusters, based on 1985 and earlier references, minus 10% random*

It is fascinating to observe that *only* on the basis of the entire reference lists, publications create a BC network with -at least up to large cluster sizes- a scale-free, power-law behaviour. We conclude from this observation that the 'affinity' of publications with other publications is only optimal in the case of complete reference lists. If one deletes a part of a publication's reference list, the publication is not anymore 'what it is', not fully characterized (remember that in the BC process a publication is as it were represented by its set of references). As specific 'affinity' leads to 'preferential attachment' (just as in molecular attachment processes), and as this preferential attachment is a strong condition for scale-free behaviour of networks, it is plausible to say that only in the case of BC networks based on the set complete of references we will find a power-law distribution function.

Removing references is a kind of imposing constraints to the publications, which are the nodes of the BC network. From earlier work we know that preferential attachment can be hindered by such constraints. Barabási and Albert (1999) [20] show the influence of constraints, demonstrating the difference between 'physical' networks (co-author networks, electrical power grids) which clearly 'suffer' from constraints, and 'virtual' networks such as the Worldwide Web where such physical constraints do not play a role (see also Dorogovtsev and Mendez 2002 [5]; Albert and Barabási 2000 [15], 2002 [8]). Amaral *et al* (2000) [26] suggest that these constraints may determine the emergence of different classes of networks, and we believe that this effect is visible in our age-dependent 'tuning' through the references.

# 4. Concluding remarks

An important aspect of real-world networks is their growth (Klemm and Eguíluz 2000b [27]). Currently we are investigating the properties of the growth of our BC publication network. This network grows simply by adding the next year of publications, which would be 2002 in our case. Updating could also be done on a weekly or monthly base. Thus, for our network dynamical evolution is explicitly available. This growth process introduces an intriguing phenomenon. A smaller, but still considerable fraction of the references of these added 2002-publications are references to 2001-publications, which are nodes in the network.

But this does not mean that a new link is created between the new 2002-publication and the 2001-publication. A link between a new node (2002-publication) and an existing node (2001-publication) is only created if a reference of the new 2002-publication is the same as a reference in the 2001-publication, for instance both publications refer to an article published in 1999. In this rather curious way, even *clustering of old, unconnected nodes* is possible. For instance, 2001-publication $p_1$ (see Section 2) has five references $r_1$, $r_2$, $r_3$, $r_4$, and $r_5$, while 2001-publication $p_4$ has none of these five references but contains other references, say, $r_6$. In the 2001-network, $p_1$ and $p_4$ are *not linked* as they do not share any reference. If in 2002 a 'new' publication $p_5$ contains one or more of the $p_1$ references, for instance $r_2$ and $r_3$, and also reference $r_6$, this new publication $p_6$ will establish in the extended 2001 + 2002 network a link with the old nodes $p_1$ and $p_2$. In other words, $p_6$ forms a BC cluster with $p_1$ and $p_2$. Notice however that a *direct link* between the old nodes $p_1$ and $p_2$ is not created.

In general, for a newly added publication it is likely that one or several of its references will bibliographically couple this new publication to older publications that already have a large number of BC co-publications. Thus, in a growing BC network the 'older' nodes increase their connectivity leading to a reinforcement of preferential attachment. For a discussion of the measurement of preferential attachment in evolving networks, see Jeong et al (2003) [28] and Dorogovtsev and Mendes (2002) [5].

Mossa *et al* (2002) [29] make a connection with our earlier work on the growth of scientific literature (van Raan 1990 [31], 2000 [30]) by considering the situation in which new nodes are not processing information from a *constant fraction* of existing nodes, but from a *constant absolute number* of nodes. In other words, as the network grows, the new nodes are only able to process information about a relatively small fraction of existing nodes. This model is plausible for networks that have grown to a very large size, for instance the scientific literature. They conclude that the above process reinforces preferential attachment, and with that, clustering which is in fact similar to fragmentation. Therefore, in our current work we investigate whether this is indeed the case for the BC network, which means, more concretely, whether the distribution function of number of references per paper does not change significantly as a function of time. Furthermore, we will study in more detail how our citation distribution model discussed above could help to construct a theoretical framework to better describe the behavior of processes taking place on networks.

*Acknowledgements*
The author thanks Peter Negenborn for his extensive data-analytical and programming work.

**Appendix**

The basic concept of our model (van Raan 2001 [23]) is the idea that scientific communication is characterized by a large number of publications that has to be divided according to attributed *status* under three assumptions: (1) the total system of scientific communication contains a limited amount of attributable status; (2) the status of a publication is represented in a significant way by the status of the journal in which it is published; and (3) the status of a journal is operationalized significantly by the way it is cited by other journals ('bibliometric' operationalization).

Our model consists of two steps. First, competition amongst scientists for 'publication status'. We argue that the underlying distribution originates from an equilibrium distribution of publications according to their 'status'. This 'status' is determined by the journal in which a publication appears and it is operationalised by the extent to which the journal is cited by other journals. Second, within their status level, scientists again have to compete with their publications (i.e., with their 'work'), in terms of getting cited ('income'). On the basis of these two basic distributions, a final one results, the distribution of citations (i.e., citing publications) over source publications.

Given these assumptions, we calculate the most probable distribution of publications over status levels. The probability of any specific distribution is proportional to the number of ways this distribution can be realized. Thus we calculate this distribution following the lines of statistical mechanics which leads us to a Boltzmann distribution of publication numbers *N* over journal status *W*:

$$N = A \cdot e^{-\alpha W} \qquad (Eq.\ A1)$$

This result (exponential distribution) is clearly supported by empirical findings, see for instance Seglen (1992) [33]. The operationalization of the journal status *W* is based on the bibliometric indicator *JCS* as discussed in Van Raan (1996) [32]. This indicator is related to the 'impact factor' of a journal but it is defined differently, in order to cover a larger time-period for citations and to take article types into account.

We rewrite the *distribution function* given in Eq. A1 as a *density function*:

$$\rho(W) = N \alpha \exp(-\alpha W), \text{ with } \int_0^\infty \rho(W)\, dW = N \qquad (Eq.\ A2)$$

We now suppose that the probability for publications to be cited *within a journal* is the probability to occupy *internal* status-levels with the same rules as discussed in the first step. Thus we find for this probability to be cited *c* times:

$$p(c) = b\exp(-b\,c), \quad \text{with } \int_0^\infty p(c)\,dc = 1 \tag{Eq. A3}$$

With help of Eq. A3 the average number of citations per publication $<c>$ can be written as

$$W = <c> = 1/b \tag{Eq. A4}$$

Given the empirical fact that our status parameter $W$ can be considered in good approximation as a continuous variable, we rewrite the probability function for the distribution of publications within a specific journal over the received citations $c$ with help of Eqs. A3 and A4:

$$p(c) = (1/W)\cdot \exp(-c/W) \tag{Eq. A5}$$

The probability that a publication in a given journal will receive a specific number of citations is then given by:

$$\rho(W, c) = \rho(W)\cdot p(c) = N\,\alpha\,\exp(-\alpha W)\,(1/W)\,\exp(-c/W) \tag{Eq. A6}$$

Finally we arrive at the distribution of all publications over citations:

$$N(c) = \rho(c) = \int_0^\infty \rho(W, c)\,dW = N\,\alpha \int_0^\infty \exp(-\alpha W - c/W)\,(1/W)\,dW \tag{Eq. A7}$$

The integral in Eq.A7 is a *modified Bessel-function of the 0-th order*, and thus we find

$$N(c) = 2\,N\,\alpha\,K_0(2\sqrt{\alpha c}) \tag{Eq. A8}$$

Empirical distribution functions of citations do *not* follow a power law for the lower numbers of citations. This is a serious problem, as most of the publications receive just a few citations. Our model solves this problem, as it fits very well with the empirical data (van Raan 2001 [23], here we find $\alpha = 0.32$). The modified Bessel distribution however approaches a power-law behaviour, particularly for the higher numbers of citations, in agreement with all observations.

Even the value for zero citations is predicted very well. We find this value by the following argument. The number of citations is by definition an integer. Thus we deal with a discrete distribution, whereas the Bessel function holds for a continuous distribution. So we approximate the $c$-values with the nearest integer, which means integration of the Bessel function. For instance: the probability for zero citations is given by the integration of $N(c)dc$ from $c = 0$ to $0.5$, i.e., the 'cumulative chance':

$$\int_0^{0.5} N(c)\,\mathrm{d}c$$

With parameter $\alpha = 0.32$ as discussed above, this integration of the Bessel function yields 0.310, and the measured (relative) number is 0.292.

With help of this two-step competition model we find that the distribution of citations over publications does not follow a power-law, but is represented by a modified Bessel function. We find a very good agreement between the outcomes of our model and empirical data.

The citation distribution process can be seen as a specific representation of a more generic process of income distribution. Thus, our two-step competition model may be of interest for the understanding of complex social and economic phenomena. For instance, the income distribution of may result from a process in which people first have to compete (with education, talent, etc.) for occupations of different 'status' in society, and, second, within these occupations for their own position in terms of salary, revenues, etc. Thus, we wonder if the famous Pareto distributions are indeed power-law distributions, or, according to a more generic form of our two-step competition model, a modified Bessel function.

*References*

[1] Redner, S. (1998). How popular is your paper? An empirical study of the citation distribution. *Eur. Phys. J. B* 4, 131-134.
[2] Vazquez, A. (2001). Statistics of citation networks. *E-print arXiv*: cond-mat/0105031.
[3] Klemm, K. and V.M. Eguíluz (2002a). Highly clustered scale-free networks. *Physical Review E*, 65, 036123.
[4] Newman, M.E.J. (2003). The structure and function of complex networks. *E-print arXiv*: cond-mat/0303516.
[5] Dorogovtsev, S.N. and J.F.F. Mendes (2002). *Advances in Physics* 51, 1079-1187.
[6] Barabási, A.-L., H. Jeong, Z. Néda, E. Ravasz, A. Schubert, T. Vicsek (2002). Evolution of the social network of scientific collaborations. *Physica A* 311, 590-614.
[7] Watts, D.J. and S.H. Strogatz (1998). Collective dynamics of 'small-world' networks. *Nature* 393, 440-442.
[8] Albert, R. and A.-L. Barabási (2002). Statistical mechanics of complex networks. *Rev. Mod. Phys*. 74, 47-97.
[9] Newman, M.E.J. (2001a). Scientific collaboration networks. I. Network construction and fundamental results. *Physical Review E*, 64, 016131.
[10] Newman, M.E.J. (2001b). Scientific collaboration networks. II. Shortest paths, weighted networks, and centrality. *Physical Review E*, 64, 016132.
[11] Newman, M.E.J. (2001c). The structure of scientific collaboration networks. *Proc. Nat. Academy of Sciences* 98, 404-409.
[12] Price, D.J. de S. (1965). Networks of scientific papers. *Science* 149, 510-515. See his ref. 3 for the papers of M.M. Kessler on bibliographic coupling.
[13] Lotka, A.J. (1926). The frequency distribution of scientific productivity. *J. Washington Acad. Sci.* 16, 317-323.
[14] Krapivsky, P.L., S. Redner and F. Leyvraz (2000). Connectivity of growing random networks. *Phys. Rev. Lett*. 85, 4629-4632.
[15] Albert, R. and A.-L. Barabási (2000). Topology of evolving networks: local events and universality. *Phys. Rev. Lett*. 85, 5234-5237.
[16] Price, D.J. de S. (1976). *J. Amer. Soc. Inform. Sci. (JASIS)* 27, 292-306.
[17] Naranan, S. (1971). Power law relations in science bibliography- a self-consistent interpretation. *J. of Documentation* 27, 83-97.
[18] Bornholdt, S. and H. Ebel (2001). World Wide Web scaling exponent from Simon's 19555 Model. *Phys. Rev. E 64*, 035104.
[19] Simon, H.A. (1955). On a Class of Skew Distribution Functions. *Biometrika* 42, 425-440.
[20] Barabási, A.-L. and R. Albert (1999). Emergence of scaling in random networks. *Science* 286, 509-512.
[21] Laherrère, J. and D. Sornette (1998). Stretched exponential distributions in nature and economy: "fat tails" with characteristic scales. *Eur. Phys. J. B* 2, 525-539.
[22] Tsallis, C. and M.P. de Albuquerque (2000). Are citations of scientific papers a case of nonextensivity? *Eur. Phys. J. B* 13, 777-780.
[23] van Raan, A.F.J. (2001). Two-Step Competition Process Leads to Quasi Power-Law Income Distributions. Application to Scientific Publication and Citation Distributions. Physica A 298 (2001) 530-536.
20

[24] Dorogovtsev, S.N. and J.F.F. Mendez (2000), Evolution of networks with aging of sites. *Phys. Rev. E 62* 1842-1845.
[25] Barabási, A.-L., R. Albert and H. Jeong (1999). Mean-field theory for scale-free random networks. *Physica A* 272, 173-187.
[26] Amaral, L.A.N., A. Scala, M. Barthélémy, and H.E. Stanley. (2000). Classes of small-world networks. *Proc. Nat. Academy of Sciences*, 97, 11149-11152.
[27] Klemm, K. and V.M. Eguíluz (2002b). Growing scale-free networks with small-world behavior. *Physical Review E*, 65, 057102.
[28] Jeong, H., Z. Néda and A.-L. Barabási (2003). Measuring preferential attachment in evolving networks. *Europhys. Lett*. 61, 567-572.
[29] Mossa, S., M. Barthélémy, H.E. Stanley and L.A.N. Amaral (2002). Truncation of power law behavior in 'scale-free' network models due to information filtering. *Phys. Rev. Lett*. 88, 138701.
[30] van Raan, A.F.J. (2000). On growth, ageing and fractal differentiation of science. *Scientometrics* 47, 347-362.
[31] van Raan, A.F.J. (1990). Fractal dimension of co-citations. *Nature* 347, 626.
[32] van Raan, A.F.J. (1996). Advanced Bibliometric Methods as Quantitative Core of Peer Review Based Evaluation and Foresight Exercises. *Scientometrics* 36, 397-420.
[33] Seglen, P.O. (1992). The skewness of science. *J. of the American Society for Information Science (JASIS)* 43, 628-638.